\documentclass[aps, prst, reprint]{revtex4-1}

\usepackage{graphicx, epsfig, array, subfigure, epstopdf, hyperref, amsmath, color}
\usepackage{MnSymbol, wasysym}
\usepackage[normalem]{ulem}

\newcommand{\dd}{\mathrm{d}}

\begin{document}

\title{Strong Meissner screening change in superconducting radio frequency cavities due to mild baking }

\author{A. Romanenko} 
\email{aroman@fnal.gov}
\author{A. Grassellino}
\author{F. Barkov}
\affiliation{Fermi National Accelerator Laboratory, Batavia, IL 60510, USA}
\author{A. Suter} 
\author{Z. Salman} 
\author{T. Prokscha} 
\affiliation{Laboratory for Muon Spin Spectroscopy, Paul Scherrer Institute, CH-5232 Villigen PSI, Switzerland} 

\date{\today}

\begin{abstract}
We investigate ``hot'' regions with anomalous high field dissipation in bulk niobium superconducting radio frequency cavities for particle accelerators by using low energy muon spin rotation (LE-$\mu$SR) on corresponding cavity cutouts. We demonstrate that superconducting properties at the hot region are well described by the non-local Pippard/BCS model for niobium in the clean limit with a London penetration depth $\lambda_\mathrm{L} = 23 \pm 2$~nm. In contrast, a cutout sample from the 120$^\circ$C baked cavity shows a much larger $\lambda > 100$~nm and a depth dependent mean free path, likely due to gradient in vacancy concentration. We suggest that these vacancies can efficiently trap hydrogen and hence prevent the formation of hydrides responsible for rf losses in hot regions.
\end{abstract}

\maketitle

Superconducting radio frequency (SRF) cavities are the key technology for future particle accelerators for high-energy physics, nuclear physics, light sources, and accelerator-driven subcritical reactors. The invention of the ``cold'' technology - where the beam accelerating structures are made of superconducting niobium instead of normal conducting copper - revolutionized accelerators, cutting the required operational power by orders of magnitude, and allowing to sustain very high electric fields of $\sim$60 MV/m at the 100\% duty factor~\cite{Padamsee_Review_SUST_2001}. 

Several decades of SRF R\&D at laboratories and universities worldwide have lead to the successful realization of niobium cavities that reliably achieve very high gradients and quality factors~\cite{Hasan_book2, Padamsee_Review_SUST_2001}. However, these structures suffer from a systematic effect of decreasing efficiency for increasing accelerating voltages. This long-standing critical problem is due to the emergence of highly dissipative ``hot'' regions on cavity surface - a phenomenon known as the \emph{high field $Q$ slope}. Extensive studies~\cite{Hasan_book2} demonstrated that all parts of the cavity surface become ``hot'' regions as soon as the local amplitude of the magnetic field reaches $B_\mathrm{rf} \gtrsim 100$~mT. For electropolished cavities an empirically found treatment - 120$^\circ$C baking for 48 hours - causes ``hot'' regions to disappear and the whole cavity surface becomes ``cold'' in a sense that there is no anomalous extra losses emerging at $B_\mathrm{rf} \gtrsim 100$~mT.

Despite intense investigations~\cite{Hasan_book2}, the nature of the hot regions and the mechanism of the $120^\circ$C baking are still subjects of debate. Understanding the cause of these losses and finding the best ways to overcome them are the keys to push SRF cavities performance and to significantly reduce costs for current and future accelerators worldwide. A way to gain this understanding is by studying the differences in microscopic superconducting properties of hot regions and non-dissipative cold regions. Precise microscopic measurements of the magnetic field profile $B(z)$ \textit{inside} superconductors recently became possible with the development of the low energy muon spin rotation technique (LE-$\mu$SR)~\cite{Morenzoni_PRL_1994, Morenzoni_JPhysCond_2004}. The unmatched sensitivity of LE-$\mu$SR was demonstrated on thick films of Nb in the clean limit, where a clear evidence for a nonlocal electromagnetic response~\cite{Suter_PRL_2004, Suter_PRB_2005} was found, a finding beyond the reach of other existing techniques. Coupled with the ideally suitable depth range of 0-130~nm and spot size of about 1~cm, LE-$\mu$SR is an ideal probe to address the problem of hot region emergence. 

In this article, we directly reveal the superconducting properties within 130~nm from the surface at the hot region of an
electropolished (EP) niobium SRF cavity and in the cold non-dissipative region from the 120$^\circ$C baked EP cavity. This is achieved by measuring $B(z)$ beneath the surface with LE-$\mu$SR: due to the Meissner effect the superconductor expels the applied external field from its interior on a length scale given by the London penetration depth $\lambda_\mathrm{L}$, a fundamental microscopic parameter of a superconductor which is directly related to the density of Cooper pairs and the electron mean free path. We demonstrate that the hot region is well described by the non-local clean limit BCS/Pippard electrodynamics with $\lambda_\mathrm{L}=23\pm2$~nm. In contrast, cutout from the 120$^\circ$C baked cavity is found to have a much larger $\lambda > 100$~nm consistent with a strongly suppressed electron mean free path. Interestingly and counter-intuitively, a much larger penetration depth -- normally indicative of a weaker, dirtier superconductor -- actually leads to a much lower high rf field dissipation. Identification of the physical mechanism for hot region mitigation suggests an alternative route by, for example, impurity doping. 

\begin{figure*}[htb]
\includegraphics[width=\linewidth]{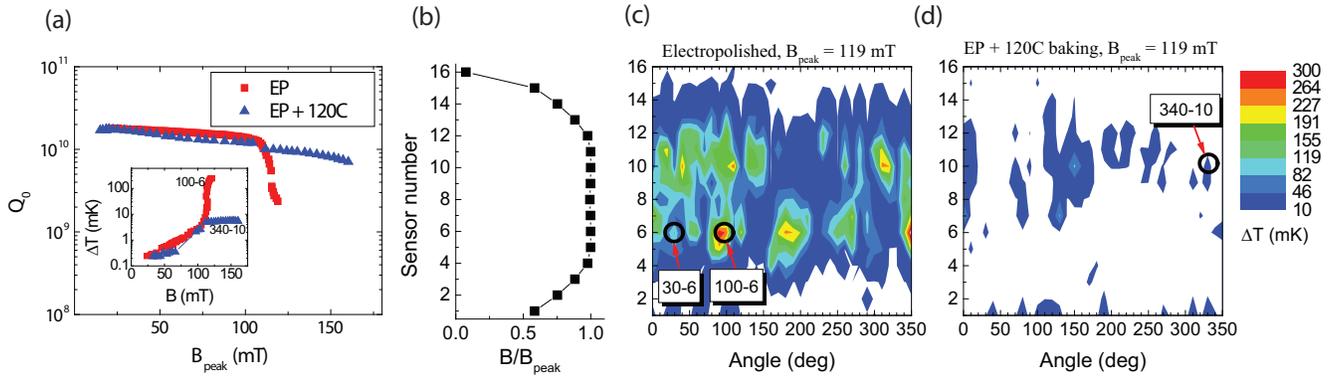}
\caption{\label{fig:RFresults}(a)
The intrinsic cavity quality factor $Q_0$ at $T=2$~K as a function of peak surface magnetic field $B_\mathrm{peak}$ on the cavity surface for the electropolished unbaked ($\blacksquare$) and 120$^\circ$C baked ($\blacktriangle$) superconducting cavities from which samples were dissected; the inset shows the rf heating $\Delta T (B)$ for cutout samples -  notice the strong correlation between the $\Delta T$ increase and  $Q_0$ degradation. (b) Calculated magnetic field distribution on the surface of the cavity at the locations of the temperature sensors. (c)-(d) ``Unfolded'' temperature maps of the outside walls at $B_\mathrm{peak}=119$~mT for: (c) unbaked electropolished cavity; (d) electropolished + 120$^\circ$C baked cavity. White areas correspond to $\Delta T<10$~mK. Locations for sample cutout are marked on the maps by circles.}
\end{figure*}

In order to obtain samples with well-characterized microwave dissipation, the best and straightforward way, although destructive, is to perform temperature mapping measurements on state-of-the-art SRF cavities followed by dissection of areas of interest from the walls. Such an approach was proven to be extremely useful in past investigations of field emission and thermal breakdown~\cite{Knobloch_Thesis}, and lately for the high field losses~\cite{Romanenko_SUST_2010, Romanenko_SUST_ERD_2011, Grassellino_PRST_muSR_2013}. For our studies we used two niobium SRF cavities of TESLA elliptical shape~\cite{TESLA_Cavities_PRST_2000} with a residual resistivity ratio RRR$\sim300$. After manufacturing both cavities were electropolished for removal of about 120~$\mu$m material using a standard solution of HF:H$_2$SO$_4$:HPO$_3$. One of the cavities was baked in vacuum at 120$^\circ$C for 48 hours as a last step. Detailed measurements of the microwave dissipation in the fundamental TM$_{010}$ mode with $f_0 = 1.3$~GHz were performed at $T=2$~K using both standard phase-lock techniques~\cite{Knobloch_Cavity_Meas_1991, HasanBook} and, independently, by a temperature mapping system similar to Ref.~\cite{Knobloch_T_Map_RSI_1994}, attached to the outside cavity walls. Such local thermometry consists of 576 Allen-Bradley carbon resistors arranged in 36 boards (equally spaced every 10$^\circ$ around the cavity rotational axis) with 16 thermometers in each board. It measures the heating of the outside cavity wall caused by the microwave dissipation on the inside surface. The local temperature increase $\Delta T$ at each thermometer location is proportional to the dissipated power on the inside wall, $\Delta T \propto P_\mathrm{diss} \propto R_\mathrm{s} (B) B^2$, thereby providing a direct measurement of the local surface resistance $R_\mathrm{s}$ since the distribution of surface magnetic field is known from numerical calculations. 

\begin{figure}[htb]
\includegraphics[width=\linewidth]{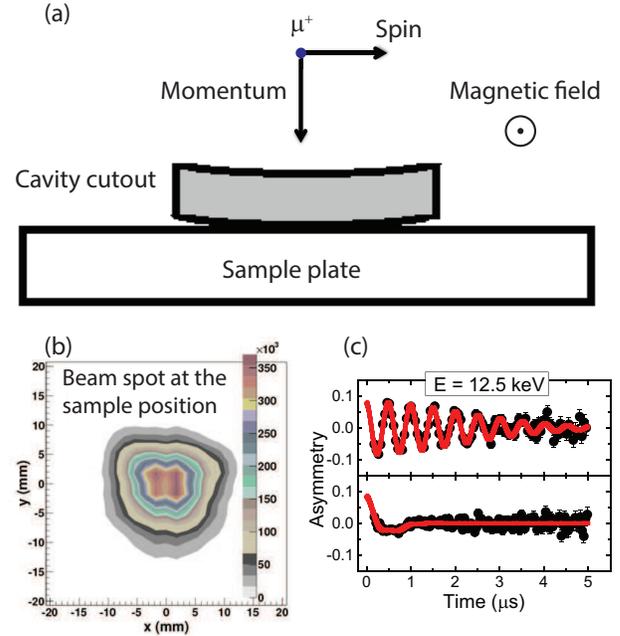}
\caption{\label{fig:LEM_schematic}(a) Schematic of the experiment; (b) measured muon flux distribution; (c) asymmetry signals $A(t)$ in normal (top) and superconducting (bottom) states of the EP sample (100-6) at muon implantation energy $E=12.5$~keV corresponding to the mean stopping depth of about 42~nm.
}
\end{figure}

Quality factors $Q_0$ of both cavities measured at $T=2$~K as a function of $B_\mathrm{peak}$ are shown in Fig.~\ref{fig:RFresults}(a). The drastic difference at high fields is a typical result of the 120$^\circ$C baking~\cite{Kneisel_SRF99}. Typical ``unfolded'' temperature maps at \mbox{$B_\mathrm{peak}=119$~ mT} are shown in Fig.~\ref{fig:RFresults}(c)-(d). A clear difference is apparent - the unbaked cavity shows a much stronger heating within the belt between sensors number 4 to 12, which corresponds spatially to the high surface magnetic field on the inside cavity surface
 [Fig.~\ref{fig:RFresults}(b)]. The surface magnetic field at different sensor locations is shown in Fig.~\ref{fig:RFresults}(b). Notice that it is very close to the peak surface magnetic field $B_\mathrm{peak}$ for sensors 4-12. Based on the rf measurements we have selected representative samples - (100-6) for the unbaked cavity, and (340-10) for the 120$^\circ$C baked cavity - which (due to 120$^\circ$C bake) have drastically different $\Delta T (B)$ as shown in the inset of Fig.~\ref{fig:RFresults}(a). Circular cutouts of $\sim$11~mm diameter were extracted from these locations using a low rotary speed milling machine with no lubricant to minimize contamination~\cite{Romanenko_thesis}. Another sample extracted from the location (30-6) in the EP unbaked cavity and subjected to buffered chemical polishing (BCP) for 20~$\mu$m material removal was used to represent unbaked BCP-treated cavities which exhibit similar $Q_0 (B_\mathrm{peak})$ behavior to the EP ones~\cite{Hasan_book2}. All three samples have been subsequently investigated with LE-$\mu$SR.

The $\mu$SR technique uses beams of 100\% spin-polarized positive muons ($\mu^+$), which serve as sensitive local magnetic probes when implanted inside a sample~\cite{muSR_book_Yaouanc}. At the $\mu$E4 beam line~\cite{Prokscha_NIM_2008} at PSI a high intensity surface muon beam with an energy of $\sim$4~MeV is moderated to ultra-low epithermal energies ($\sim$15~eV) in a cryogenically condensed solid Ar film deposited on a 10-K-cold Ag foil. These epithermal muons are subsequently accelerated by electrostatic fields to energies $E \leq 30$~keV, corresponding to
implantation depths up to $\sim$140~nm in Nb [see Supplementary Material]. The schematic of the experimental arrangement and measured muon flux distribution on the sample are shown in Fig.~\ref{fig:LEM_schematic}(a)-(b). Upon implantation, the muon precesses in the local magnetic field at its stopping site. The precession frequency is proportional to the magnetic field and is measured by detecting the anisotropic muon decay (lifetime $\tau_{\mu} = $~2.2~$\mu$s): the decay positrons are preferentially emitted in the direction of the $\mu^+$ spin, which allows to monitor the time evolution of the muon spin by registering the positrons in detectors surrounding the sample. The number of positron events at each of the detectors is described by the following form:
\begin{equation}
N(t) = N_0 \exp(-t/\tau_\mu) \left[1 + A(t)\right] + N_{\rm bkg},
\end{equation}
\noindent where $A(t)=A_0P(t)$ describes the time evolution of the muon ensemble polarization $P(t)$, and $A_0$ is the experimental decay asymmetry. $N_{\rm bkg}$ is a time-independent uncorrelated background. The asymmetry $A(t)$ is given by averaging over the muon stopping distribution $n(z, E)$:
\begin{equation}
A(t) = A_0\, \exp\left[-\frac{(\sigma t)^2}{2}\right] \int n(z, E) 
\mathrm{cos}\left[ \gamma B(z) t + \phi \right]\, \dd z,
\end{equation}
where $\gamma = 2 \pi \times 135.54$~MHz/T is the muon gyromagnetic ratio, $\phi$ is the detector phase, and $\sigma$ is a Gaussian depolarization rate, reflecting the dipolar broadening due to nuclear spins.

In the first set of experiments we performed zero-field cooling to $T=3$~K and then applied a magnetic field, $B_{\rm a}$, parallel to the sample surface and transverse to the muon spin [see Fig.~\ref{fig:LEM_schematic}(a)]. The magnitude of $B_\mathrm{a}$ was confirmed in each case by performing a run in the normal state at $T=10$~K above the transition temperature of niobium ($T_\mathrm{c}=9.25$~K) where the Meissner effect is absent. Muon implantation energies of $3.3\leq E \leq 25.3$~keV were used. Corresponding implantation profiles simulated using the computer code TRIM.SP~\cite{Eckstein_TRIM_SP, Morenzoni_NIM_2002} can be found in [Supplementary Material]. For the simulations, niobium oxide (Nb$_2$O$_5$) of 5~nm thickness was assumed as the topmost layer~\cite{Grundner_Oxides_1979}. Systematic uncertainty in these simulations is estimated to be of order 2\%, which translates into $\leq$2~nm of the mean depth uncertainty.

Several million decay positrons were collected for each muon energy. Examples of asymmetry signals $A(t)$ obtained on the same sample (100-6) at $B_\mathrm{a}=15$~mT, $E=12.5$~keV in the normal and superconducting states are shown in Fig.~\ref{fig:LEM_schematic}(c). The Meissner effect becomes manifested in the reduction of the precession frequency and the heavily damped $A(t)$ caused by the broad field distribution in the stopping range of the muons.

\begin{figure}[htb]
\includegraphics[width=\linewidth]{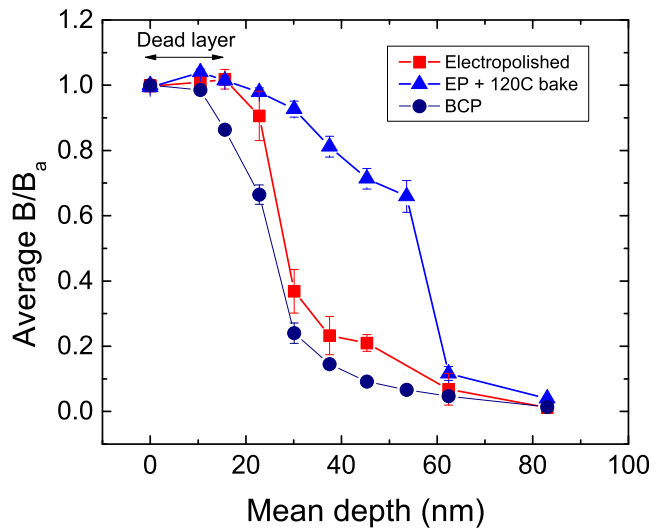}
\caption{\label{fig:B_z_all}Magnetic field profiles $\langle B \rangle (\langle z \rangle)$ at 
 $B_\mathrm{a}=25$~mT using the Gaussian model in the EP cavity cutout (100-6) ($\blacksquare$) and EP+120$^\circ$C cavity cutout (340-10) ($\blacktriangle$), and at $B_\mathrm{a}=28$~mT for the BCP treated sample (30-6) ($\CIRCLE$). Lines are guides to the eye. 
}
\end{figure}

All data were analyzed using the program \emph{musrfit}~\cite{Suter_Wojek_PhysProcedia_2012}.
We used two fit models: a simple Gaussian model, and a numerical time-domain model based on the non-local Pippard/BCS model. For comparing $B(z)$ between the samples, we first use the well-established Gaussian approximation~\cite{Suter_PRB_2005}:
\begin{equation}
  A(t) = A_0\, \exp\left[-\frac{(\sigma_{\rm SC} t)^2}{2}\right] \cos(\gamma_\mu B_{\rm G} t + \phi),
\end{equation}
\noindent where $B_{\rm G}$ is in \emph{very} good approximation equal to $\langle B \rangle$ which is given by
\begin{equation}
\langle B \rangle = \int_0^{\infty} B(z)\,n(z,E)\, \dd z.
\end{equation}
The screened magnetic field $B_{\rm G}$ as a function of the mean muon stopping depth
\begin{equation}
 \langle z \rangle = \int_0^{\infty} z\, n(z,E)\, \dd z
\end{equation} 
is presented in Fig.~\ref{fig:B_z_all} for all three samples.

A drastic difference is apparent in the Meissner screening brought about by 120$^\circ$C bake, which correlates with the strong difference in surface resistance. It is remarkable that the seemingly ``poorer'' superconductor (340-10) is actually less dissipative under microwave fields [see Fig.~\ref{fig:RFresults}(a)]. The field profiles $B(z)$ at all investigated fields - $B_\mathrm{a} = 5, 15, 25$~mT for (100-6), and $B_\mathrm{a} = 10, 25$~mT for (340-10) - are very similar to the ones in Fig.~\ref{fig:B_z_all}.  

\begin{figure}[htb]
 \includegraphics[width=\linewidth]{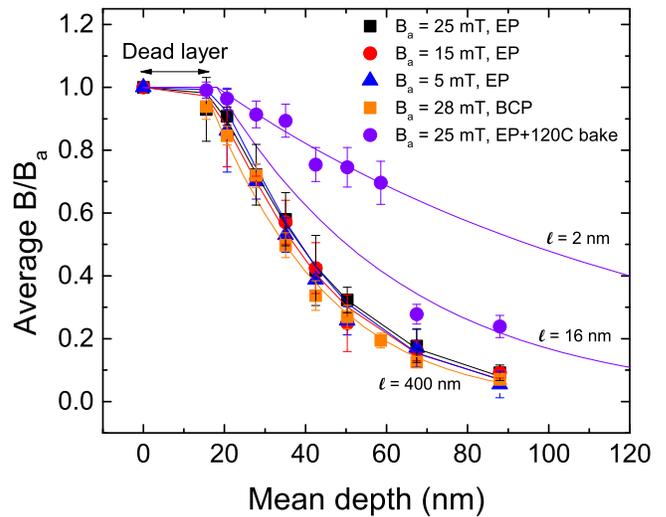}
\caption{\label{fig:Hot_B_z}Average normalized field vs. mean muon stopping depth. Individual points correspond to the Pippard fits based on single energy measurements only. Solid lines for unbaked EP (100-6) and BCP (30-6) samples show global fits based on the Pippard/BCS model. Solid lines for the EP+120$^\circ$C cavity cutout (340-10) show Pippard/BCS model calculations for different values of the electron mean free path $\ell$.
}
\end{figure}

Previous studies on thick niobium films in the clean limit~\cite{Suter_PRL_2004, *Suter_PRB_2005} showed that a very good description of the experimental data is provided by Pippard/BCS non-local models with the renormalization of $\lambda_\mathrm{L} \to \lambda_\mathrm{L}/\sqrt{Z} $ and of the coherence length $\xi_0 \to \xi_0 Z$ where $Z \approx 2.1$~\cite{Carbotte_RMP_1990}. Since earlier SRF cavity experiments~\cite{Saito_Kneisel_SRF_1999} showed that electropolished and buffered chemical polished niobium are also in the clean limit ($\ell \gg \xi_0$, $\ell$ being the electron mean free path), a similar behavior is expected for our samples. Following the methodology in~\cite{Kiefl_PRB_2010} we fitted our time-domain data of the EP unbaked (100-6) and BCP unbaked (30-6) samples using the model field distribution: 
\begin{equation}
B(z) = 
\begin{cases}
B_\mathrm{a}, & \text{if } z \leq d \\
B_\mathrm{P}(z) & \text{if } z > d,
\end{cases}
\end{equation}
with $d$ the thickness of a ``dead'' layer where the magnetic field is not screened, and $B_\mathrm{P}(z)$ the Pippard/BCS model prediction for an ideally flat semi-infinite superconductor evaluated numerically. The presence of a ``dead'' layer is a general feature observed in many LE-$\mu$SR studies. While the true origin of $d$ has not yet been firmly established, a plausible explanation is based on surface roughness~\cite{Jackson_PRL_2000} which is also theoretically supported~\cite{Dead_Layer_JEngMath_2013}. We used the literature values $\xi_0=39$~nm and $\ell=400$~nm as fixed input parameters, while $d$ and $\lambda_\mathrm{L}$ were allowed to vary. All fits were excellent ($\chi^2/\mathrm{NDF} \simeq 1.05$), and resulting individual energy fits (points) and global fits (lines) at all applied fields are shown in Fig.~\ref{fig:Hot_B_z}.  
Values of $\lambda_\mathrm{L}$ extracted from global fits at different fields are shown in Table~\ref{Lambdas}.
\begin{table}[htb]
\caption{\label{Lambdas}Values of $\lambda_\mathrm{L}$ from global fits.}
\begin{ruledtabular}
\begin{tabular}{cccc}
Cavity/sample & Applied magnetic field (mT) & $\lambda_\mathrm{L}$ (nm)\\ 
\hline
EP (100-6) & 5 & 21.8(4) \\
                     & 15 & 23.5(4) \\
                     & 25 & 24.4(5) \\
\hline                      
BCP (30-6) & 28 & 24.8(4) \\
\end{tabular}
\end{ruledtabular}
\end{table}%

Either a weakly field-dependent or a constant $\lambda_\mathrm{L}$ is consistent with our data for the electropolished unbaked sample (100-6). Assuming no field dependence, our fits give $\lambda_\mathrm{L} = 23 \pm 2$~nm and $d = 18 \pm 2 $~nm. Corresponding values for the BCP (30-6) sample are $\lambda_\mathrm{L} = 25 \pm 2$~nm and $d = 15 \pm 1 $~nm. Both EP (100-6) and BCP (30-6) samples have $\lambda_\mathrm{L}$, which are reasonably close to the thick Nb film value $\lambda_\mathrm{L} = 27 \pm 3$~nm~\cite{Suter_PRL_2004, *Suter_PRB_2005}, and are somewhat lower than that from bulk magnetometry measurements~\cite{Casalbuoni_NIM_2005} $\lambda_\mathrm{L} = 46 \pm 2$~nm. However, if the latter is ``renormalized'' to take into account non-weak coupling in niobium ($\lambda_\mathrm{L} \to \lambda_\mathrm{L}/\sqrt{Z}$, $Z \approx 2.1$) we get $\lambda_\mathrm{L} = 32 \pm 2$~nm closer to our values.

Unlike data for the (100-6) hot region sample and (30-6) BCP sample, data for the (340-10) sample from the baked cavity could not be described by the Pippard/BCS model with a single mean free path $\ell$. In this case we had to follow a different procedure: values of $\lambda_\mathrm{L}$ and $d$ were fixed to those of the unbaked sample (100-6), and $\ell$ was used as a free parameter. The assumption of the same $\lambda_\mathrm{L}$ is reasonable since the same bulk niobium is used for both samples. The assumption of the same $d$ is also reasonable since samples have the same nanoroughness of the EP-treated surface and further support is provided by the Gaussian model fits [see Fig.~\ref{fig:B_z_all}]. The results of the Pippard fitting procedure (individual points) are shown in Fig.~\ref{fig:Hot_B_z} along with the calculated depth profiles for different values of $\ell$ (solid lines). Data indicates a depth-dependent mean free path with the lower value ($\ell$=2-4~nm) in the first $\sim$60~nm followed by an increased $\ell \geq 16$~nm at larger depths.  

Previous microwave cavity studies~\cite{Kneisel_SRF99, Lilje_NIM_2004, Romanenko_HF_PRST_2013} suggested that one of the effects of the 120$^\circ$C baking on niobium may be a significant decrease of $\ell$ in the first $\sim$20-30~nm from the surface causing a crossover from clean ($\ell \gg \xi_0$) to dirty ($\ell < \xi_0$) limit. We observe a strong increase of the penetration depth in the (340-10) baked cavity sample, which can be consistently described by such a mean free path suppression. Our findings are also qualitatively consistent with measurements of $B_\mathrm{c3}$ reported in~\cite{Casalbuoni_NIM_2005} where it was also found that 120$^\circ$C baking of EP samples leads to an increase of surface $B_\mathrm{c2}$ indicating ``dirtier'' material. A strongly suppressed surface $\ell$ is also in line with the increased surface pinning reported in Ref.~\cite{Grassellino_PRST_muSR_2013}.

Small nanoscale hydrides have been lately proposed to be the cause of the anomalous high field dissipation~\cite{Romanenko_SUST_Proximity_2013}, and the 120$^\circ$C baking effect attributed to the injection of vacancies~\cite{Romanenko_VEPAS_APL_2013}. Preliminary structural investigations~\cite{Tao_JAP_2013, Barkov_JAP_Hydrides_2013} are also consistent with this picture. Within this model hydrides remain superconducting by proximity effect up to the high field dissipation onset ($B_\mathrm{rf}\sim$100~mT), which is consistent with the clean limit Pippard/BCS description at $B\leq25$~mT of LE-$\mu$SR data for the hot region (100-6). If hydride precipitation is suppressed by the $120^\circ$C baking then $\ell$ remains low upon cooldown to $T \leq$3~K, which is in agreement with our findings on the baked cavity sample (340-10). Thus our results can be explained by the assumption that efficiently trapping hydrogen is the mechanism to eliminate the formation of nanoscale hydrides (hot regions). These vacancies introduced by the $120^\circ$C baking may also explain the depth dependent mean free path after baking.

Finally, it is worth mentioning that we have also performed zero-field measurements on all samples to search for any near-surface magnetic impurities motivated by recent SQUID~\cite{Casalbuoni_NIM_2005} and point contact tunneling~\cite{Proslier_PCT_APL_2008} results. However, we found no evidence for surface magnetism and no difference between the two samples.

In conclusion, we have directly measured and compared the magnetic penetration depth in hot (highly dissipative at high rf fields $B_\mathrm{rf} \gtrsim 100$~mT) regions of bulk niobium SRF cavities for particle acceleration with that in non-dissipative regions obtained by 120$^\circ$C baking. For the hot region Meissner screening at $B \leq 25$~mT is well described quantitatively by Pippard/BCS non-local electrodynamics with $\lambda_\mathrm{L} = 23 \pm 2$~nm. For the 120$^\circ$C baked cavity cutout the magnetic field penetrates much deeper ($\lambda>100$~nm) and the decay is well described by a depth-dependent electron mean free path in the range of $2 \lesssim \ell \lesssim 16$~nm. We propose that vacancies introduced by the $120^\circ$C baking may prevent hydride precipitation responsible for strong rf losses in hot regions, while also leading to the depth gradient in the electron mean free path.  Our findings suggest impurity doping of the surface layer as a possible alternative route for hot region mitigation and further SRF cavity improvement. 

Fermilab is operated by Fermi Research Alliance, LLC under Contract No. De-AC02-07CH11359 with the United 
States Department of Energy. A.R. and F.B. were partially supported by the U.S. DOE Office of Nuclear Physics.

%

\end{document}